\documentclass{appolb}
\usepackage[utf8]{inputenc}
\usepackage{epsf}
\usepackage{latexsym,amssymb,euscript}
\usepackage[dvips]{graphicx}
\usepackage[numbers,sort&compress]{natbib}
\usepackage{amsmath}
\usepackage{nicefrac}
\usepackage{slashed}
\usepackage{booktabs}
\usepackage{hyperref}
\usepackage{braket}
\usepackage{chngcntr}
\usepackage{bbold}
\usepackage{graphics}
\usepackage{graphicx}
\usepackage{mciteplus}
\usepackage{caption}
\usepackage{subcaption}
\usepackage{pdfpages}
\usepackage[titletoc]{appendix}
\graphicspath{{./figures/}}
\hypersetup{
 linktocpage = true,
 urlcolor = black,
 colorlinks = true,
 linkcolor = urlblue,
 anchorcolor = urlblue,
 citecolor = urlblue,
 pdfstartview = {XYZ null null 1.25} 
           }
\usepackage{pstricks}
\usepackage{color}
\usepackage{xcolor}
\definecolor{urlblue}{rgb}{0.2,0.4,0.7}
\definecolor{citegreen}{rgb}{0,0.4,0.2}
\definecolor{linkred}{rgb}{0.9,0.2,0.1}
\usepackage{float}
\definecolor{orcidlogocol}{HTML}{A6CE39}
\usepackage{fancyhdr}
\usepackage[utf8]{inputenc}
\usepackage[normalem]{ulem}
\usepackage{mathtools}
\usepackage{setspace}

\newcommand{\drv}{{\rm d}}

\newcommand{\LQCD}{\Lambda_{\rm QCD}}

\newcommand{\Jps}{J/\psi}
\newcommand{\Yps}{\Upsilon}

\newcommand{{\HFNRevo}}{\tt HF-NRevo}

\begin{document}
\title{On the quarkonium-in-jet collinear fragmentation \\ at moderate-to-large transverse momentum%
\thanks{Presented at ``Diffraction and Low-$x$ 2024'', Trabia (Palermo, Italy), September 8-14, 2024.}
}
\author{Francesco Giovanni Celiberto
\address{Universidad de Alcal\'a (UAH), E-28805 Alcal\'a de Henares, Madrid, Spain}
}
\maketitle

\begin{abstract}
We report progress on the Heavy-Flavor Non-Relativistic Evolution ({\HFNRevo}) setup, a novel methodology to address quarkonium formation within the fragmentation approximation. 
Our study sheds light on the moderate to large transverse-momentum sector, where the leading-twist collinear fragmentation of a single parton prevails over the higher-twist fragmentation from a constituent heavy-quark pair produced in the hard scattering.
As for the initial energy-scale inputs, we rely on nonrelativistic next-to-leading calculations for all the parton-to-quarkonia fragmentation channels. 
Preliminary sets of variable-flavor number-scheme (VFNS) fragmentation functions, named {\tt NRFF1.0}, are built via an evolution-threshold enhanced DGLAP scheme.
Taking {\tt NRFF1.0} as a starting point, we use {\HFNRevo} to address the collinear fragmentation of quarkonia inside jets.
\end{abstract}

\vspace{-0.25cm}

\section{Hors d'{\oe}uvre}
\label{sec:introduction}

Studies on hadrons with open or hidden heavy flavors are crucial to understand fundamental interactions. 
Heavy quarks, due to their potential interactions with beyond-Standard-Model particles, serve as essential probes in the quest for New Physics. 
Furthermore, their masses well above the perturbative QCD threshold make them ideal candidates for precision tests of strong interaction.
The study of quarkonia, the ``hydrogen atoms'' of QCD~\cite{Pineda:2011dg}, offers a powerful means to uncover fundamental aspects of the strong force. 
Quarkonia provide benchmarks for key areas of particle physics, from high-precision investigations of perturbative QCD to explorations of the proton's internal structure.
Decays of $S$-wave bottomonia permit precise $\alpha_s$ determinations~\cite{Brambilla:2007cz,Proceedings:2019pra}, while their forward emissions test gluon PDFs at small $x$~\cite{Altarelli:1998gn_alt,Candido:2020yat_alt,Collins:2021vke_alt,Candido:2023ujx_alt}. 
Moreover, they contribute to the 3D imaging of the proton at small $x$~\cite{Garcia:2019tne,Celiberto:2018muu,Bolognino:2018rhb,Bolognino:2021niq,Celiberto:2019slj,Silvetti:2022hyc,Kang:2023doo} and moderate $x$~\cite{Boer:2015pni,Lansberg:2017dzg,DAlesio:2020eqo,Bacchetta:2020vty,Bacchetta:2024fci,Celiberto:2021zww}. 
Additionally, studies of $\Jps$ plus charm-jet photoproduction at the EIC~\cite{Flore:2020jau} enable direct measurements of the valence intrinsic-charm PDF in the proton~\cite{Ball:2022qks,Guzzi:2022rca,NNPDF:2023tyk_alt}.
The theoretical descriptions of quarkonium hadronization remain challenging. 
Despite various models, no framework fully accounts for experimental data. 
Non-Relativistic QCD (NRQCD)~\cite{Caswell:1985ui,Bodwin:1994jh} addresses this by including all Fock states of quarkonia, expanded in powers of $\alpha_s$ and $v$, the $(Q\bar{Q})$ pair’s relative velocity. 
Cross sections are expressed as sums of perturbative Short-Distance Coefficients (SDCs) and nonperturbative Long-Distance Matrix Elements (LDMEs). 
For parallels with FFs of open heavy-flavored particles, see~\cite{Mele:1990cw,Cacciari:1993mq_alt,Cacciari:2012ny,Helenius:2018uul,Helenius:2023wkn,Czakon:2021ohs,Czakon:2022pyz,Generet:2023vte,Ghira:2023bxr_alt,Bonino:2023icn_alt,Cacciari:2024kaa}.
NRQCD allows us to model quarkonium formation both at low transverse momenta, where the short-distance $(Q\bar{Q})$ pair production dominates, and at moderate-to-high transverse masses, where the single-parton fragmentation becomes significant~\cite{Cacciari:1994dr_alt}.
We study collinear fragmentation to pseudoscalar and vector quarkonia in the color-singlet (CS) state using a preliminary version of the {\tt NRFF1.0} FF sets. 
These are built within the {\HFNRevo} framework~\cite{Celiberto:2024mex_article,Celiberto:2024bxu}, which incorporates next-to-leading-order (NLO) NRQCD initial-scale inputs, DGLAP evolution, and Missing Higher-Order Uncertainties (MHOUs) through Monte-Carlo-like methods~\cite{Forte:2002fg}.

\section{Quarkonium collinear fragmentation from {\HFNRevo}}
\label{sec:HFNrevo}

\begin{figure*}[!t]
\centering

   \includegraphics[scale=0.325,clip]{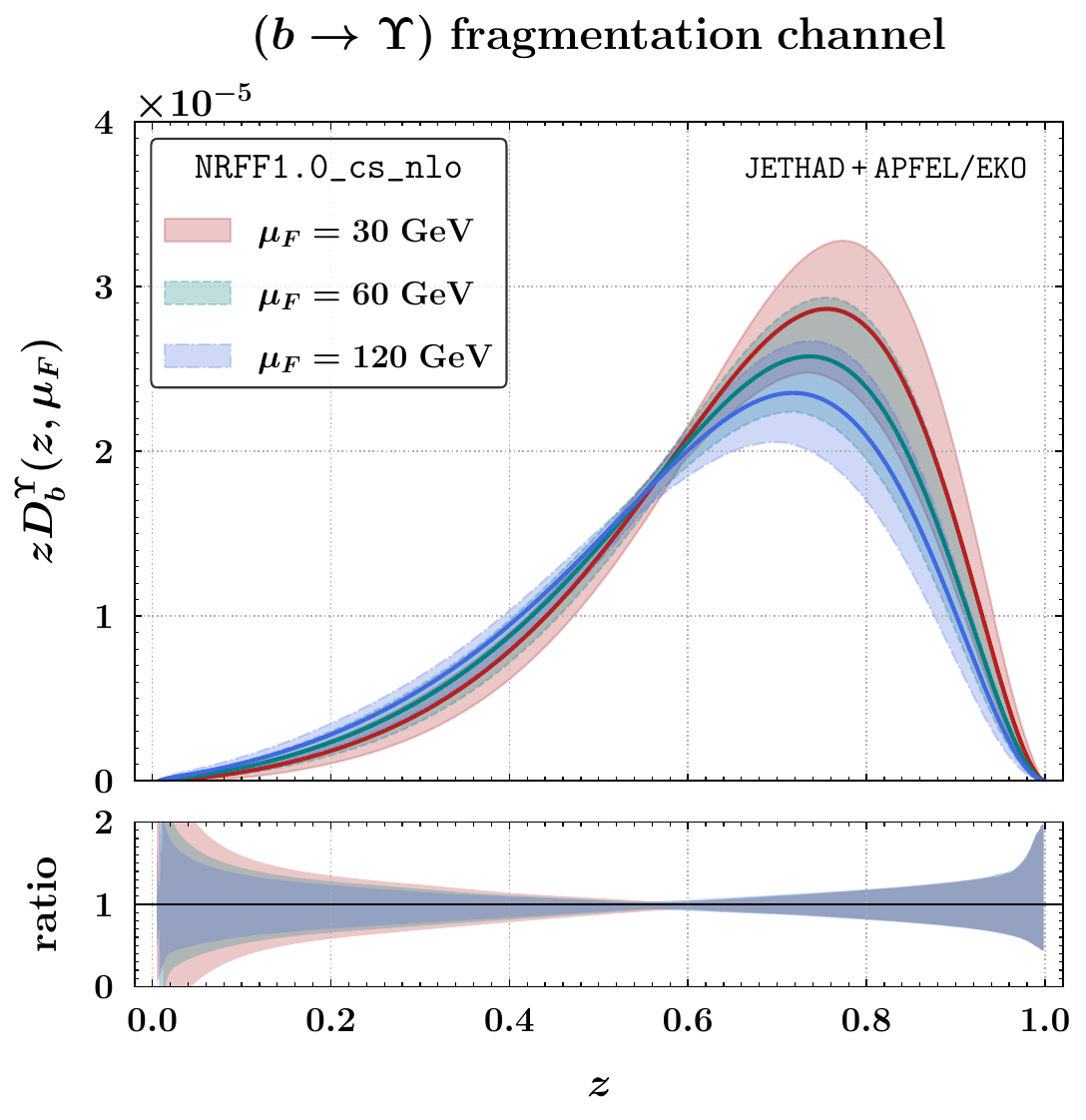}
   \hspace{0.20cm}
   \includegraphics[scale=0.325,clip]{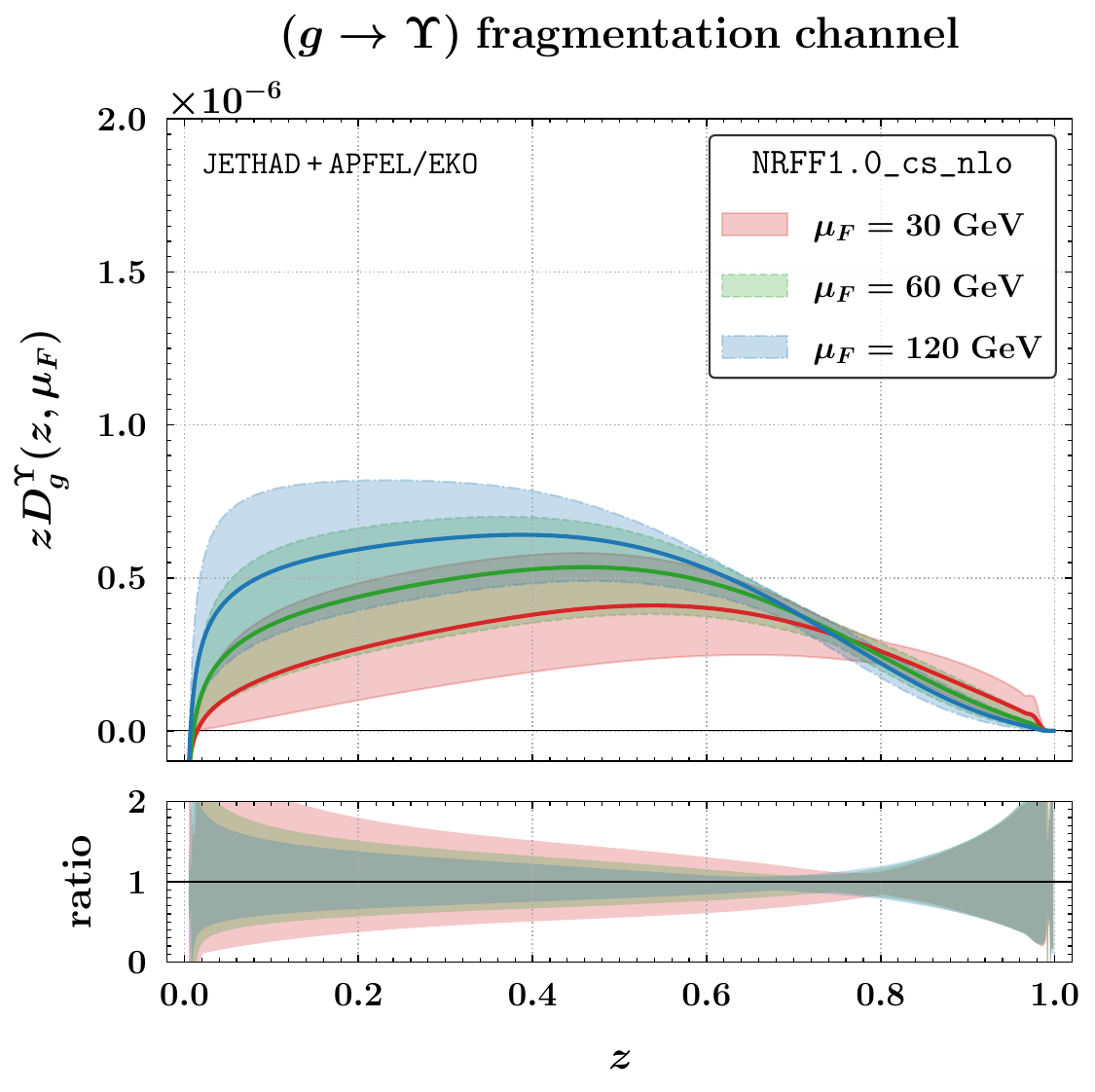}

\caption{Preliminary results for $(b,g \to \Yps)$ {\tt NRFF1.0} CS fragmentation.
}

\label{fig:FFs}
\end{figure*}

Since the masses of constituent heavy quarks are well above $\LQCD$, quarkonium FFs at their initial scale are expected to incorporate perturbative inputs. 
This necessitates a consistent use of collinear factorization. 
To address this, we propose a novel methodology named {\HFNRevo}~\cite{Celiberto:2024mex_article,Celiberto:2024bxu}, built on three pillars: \emph{interpretation}, \emph{evolution}, and \emph{uncertainties}.
The interpretation deciphers the short-distance formation at low transverse momentum as a two-parton fragmentation in a Fixed-Flavor Number Scheme (FFNS), enabling subsequent FFNS-to-VFNS matching~\cite{Kang:2014tta}. 
This is supported by distinct singularity patterns observed in the matching tails of shape functions~\cite{Echevarria:2019ynx} and 3D FFs~\cite{Boer:2023zit}.  
In {\HFNRevo}, the DGLAP evolution of quarkonium FFs proceeds in two stages. 
First, an \emph{expanded} and \emph{decoupled} evolution ({\tt EDevo}), performed symbolically using {\tt symJETHAD}~\cite{Celiberto:2017ius,Celiberto:2020wpk,Celiberto:2022rfj,Celiberto:2022kxx,Celiberto:2020tmb,Bolognino:2021mrc,Celiberto:2021dzy,Celiberto:2021fdp,Celiberto:2022zdg,Celiberto:2022gji}, accounts for the thresholds of all parton species. 
Then, the numerical \emph{all-order} evolution ({\tt AOevo}) takes over.  
Finally, MHOUs from variations in DGLAP-evolution thresholds are quantified. 
This involves a simultaneous scan of $\mu_F$ and $\mu_R$ scales in the initial FF inputs, varying them by a factor of $1/2$ to $2$, aligning with PDF analyses using theory-covariance matrices~\cite{Harland-Lang:2018bxd_alt,NNPDF:2024dpb} or the {\tt MCscales} method~\cite{Kassabov:2022orn_alt}.  
For simplicity, we present here two fragmentation channels.
Left and right plots of Fig.~\ref{fig:FFs} respectively show $(b \to \Yps)$ and $(g \to \Yps)$ {\tt NRFF1.0} NLO FFs for $\mu_F$ values ranging from 30 to 120~GeV.

\section{Quarkonium-in-jet fragmenting functions}
\label{sec:onium_in_jet}

Jet substructure measurements have recently gained prominence as a powerful sounds for core nature of the strong force. 
Beyond that, they present exciting opportunities for uncovering physics beyond the Standard Model~\cite{Altheimer:2013yza,Adams:2015hiv}. 
Advancing our theoretical comprehension of QCD dynamics by unveiling the substructure of (heavy-flavored) jets is of paramount importance~\cite{Procura:2009vm_alt,Bauer:2013bza,Chien:2015ctp,Maltoni:2016ays_alt,Kang:2017glf,Metodiev:2018ftz_alt,Marzani:2019hun_article,Kasieczka:2020nyd,Nachman:2022emq,Dhani:2024gtx}. 
In this context, relevant observables are the ones sensitive to the detection of a specific hadron identified inside a jet.
From a collinear-factorization perspective, the production mechanism of a \emph{hadron-in-jet} system is known as Semi-Inclusive Fragmenting Jet Functions (SIFJFs).
At leading power, the expression for the SIFJF of a parton $i$ fragmenting into an identified quarkonium ${\cal Q}$ inside a jet reads~\cite{Kang:2017yde}
\begin{equation}
 \label{eq:SIFJF}
 {\cal F}_i^{\cal Q}(z, z_{\cal Q}, \mu_F, {\cal R}_{\cal J}) \, = \,
 \sum_j \int_{z_{\cal Q}}^1 \frac{\drv \zeta}{\zeta} \,
 {\cal S}(z, z_{\cal Q}/\zeta, \mu_F, {\cal R}_{\cal J}) \,
 D_j^{\cal Q}(\zeta, \mu_F) \;.
\end{equation}
Here $D_j^{\cal Q}(\zeta, \mu_F)$ represents the standard $(j \to {\cal Q})$ FF channel.
Conversely, ${\cal S}(z, z_{\cal Q}/\zeta, \mu_F, {\cal R}_{\cal J})$ are the so-called collinear fragmenting jet coefficients~\cite{Baumgart:2014upa}, known at NLO for anti-$\kappa_T$ and cone jet algorithms~\cite{Kang:2016ehg_alt}.
We note that the SIFJF of Eq.~\eqref{eq:SIFJF} depends on the jet radius, ${\cal R}_{\cal J}$, and on two light-cone variables: $z$, corresponding to the energy fraction of the fragmenting parton $i$ carried by the jet, and $z_{\cal Q}$, standing for the energy fraction of the jet carried by the quarkonium ${\cal Q}$.
The next step is renormalizing the SIFJF and matching it onto the standard quarkonium FFs.
In this way, one gets the following schematic, pocket formula for the DGLAP-evolved SIFJF~\cite{Kang:2016ehg_alt,Kang:2017yde}
\begin{equation}
 \label{eq:SIFJF_evo}
 D_j^{\cal Q}(\zeta, \mu_0)
  \quad \Longrightarrow \quad
 D_j^{\cal Q}(\zeta, \mu_{\cal M})
  \quad \Longrightarrow \quad
 {\cal F}_i^{\cal Q}(z, z_{\cal Q}, \mu_F, {\cal R}_{\cal J})
 \;.
\end{equation}
Equation~\eqref{eq:SIFJF_evo} tells us that the time-like DGLAP evolution for our SIFJF proceeds through two steps.
First, we evolve the standard quarkonium FFs from the lowest scale, $\mu_0$, to the \emph{matching} value, $\mu_{\cal M}$.
Since the choice of $\mu_{\cal M}$ is arbitrary, we can set $\mu_{\cal M} \approx Q \, {\cal R}_{\cal J}$, with $Q$ the process-typical hard scale.
This minimizes logarithmic terms in the matching coefficients.
Then, from $\mu_{\cal M}$ onward, we evolve the whole SIFJF via another time-like DGLAP step, thus resumming logarithms of the form $\ln(\mu_{\cal M}) \sim \ln({\cal R}_{\cal J})$.
Therefore, the energy resummation \emph{de facto} translates into a jet radius resummation.
This paves the way towards precision studies of jet substructure \emph{via} resummation-sensitive observables, such as \emph{jet angularities}~\cite{Luisoni:2015xha,Caletti:2021oor,Reichelt:2021svh}.

\section{Towards {\tt NRFF1.0} and {\tt NRFF1.0jet}}
\label{sec:conclusions}

Using {\HFNRevo}, we developed a preliminary version of the {\tt NRFF1.0} collinear FF sets for quarkonia. 
They incorporate NLO NRQCD-based CS initial-scale inputs across all parton channels. 
A consistent DGLAP framework was established to manage evolution thresholds, and MHOUs were quantified through a Monte-Carlo replica-like approach.
The {\tt NRFF1.0} FFs will supersede the {\tt ZCW19}$^+$ and {\tt ZCFW22} functions currently employed in studies of vector quarkonia~\cite{Celiberto:2022dyf_alt,Celiberto:2023fzz} and charmed $B$ mesons~\cite{Celiberto:2022keu,Celiberto:2024omj}. 
They will play a key role in advancing quarkonium physics at the HL-LHC~\cite{Chapon:2020heu_alt,Amoroso:2022eow}, the EIC~\cite{AbdulKhalek:2021gbh,Khalek:2022bzd,Abir:2023fpo}, and new-generation lepton colliders~\cite{AlexanderAryshev:2022pkx}. 
Furthermore, they will serve as critical benchmarks for AI-driven analyses and extractions~\cite{Allaire:2023fgp,Hekhorn:2024jrj_article,Hammou:2023heg,Costantini:2024xae,Hammou:2024cwu_article}.
We aim to extend the {\HFNRevo} framework to include quarkonium-in-jet fragmentation. The future release of {\tt NRFF1.0jet} SIFJFs will offer us a novel and complementary tool for exploring the substructure of heavy-flavored jets, with promising applications in analyzing quarkonium-in-jet angularities.
Applying {\HFNRevo} to exotic hadrons is underway~\cite{Celiberto:2023rzw_alt,Celiberto:2024mrq,Celiberto:2024mab_alt}.

\section*{Acknowledgments}
\label{sec:acknowledgments}

This work is supported by the Atracci\'on de Talento Grant n. 2022-T1/TIC-24176 (Madrid, Spain).
We thank Hongxi~Xing for fruitful conversations on the production mechanism of quarkonia in jets. 

\vspace{-0.05cm}
\begingroup
\setstretch{0.6}
\bibliographystyle{apsrev}
\bibliography{bibliography}

\begin{thebibliography}{93}
\expandafter\ifx\csname natexlab\endcsname\relax\def\natexlab#1{#1}\fi
\expandafter\ifx\csname bibnamefont\endcsname\relax
  \def\bibnamefont#1{#1}\fi
\expandafter\ifx\csname bibfnamefont\endcsname\relax
  \def\bibfnamefont#1{#1}\fi
\expandafter\ifx\csname citenamefont\endcsname\relax
  \def\citenamefont#1{#1}\fi
\expandafter\ifx\csname url\endcsname\relax
  \def\url#1{\texttt{#1}}\fi
\expandafter\ifx\csname urlprefix\endcsname\relax\def\urlprefix{URL }\fi
\providecommand{\bibinfo}[2]{#2}
\providecommand{\eprint}[2][]{\url{#2}}

\bibitem[{\citenamefont{Pineda}(2012)}]{Pineda:2011dg}
\bibinfo{author}{\bibfnamefont{A.}~\bibnamefont{Pineda}}, \bibinfo{journal}{Prog. Part. Nucl. Phys.} \textbf{\bibinfo{volume}{67}}, \bibinfo{pages}{735} (\bibinfo{year}{2012}), \eprint{1111.0165}.

\bibitem[{\citenamefont{Brambilla et~al.}(2007)}]{Brambilla:2007cz}
\bibinfo{author}{\bibfnamefont{N.}~\bibnamefont{Brambilla}} \bibnamefont{et~al.}, \bibinfo{journal}{Phys. Rev. D} \textbf{\bibinfo{volume}{75}}, \bibinfo{pages}{074014} (\bibinfo{year}{2007}), \eprint{hep-ph/0702079}.

\bibitem[{\citenamefont{d'Enterria et~al.}(2019)}]{Proceedings:2019pra}
\bibinfo{author}{\bibfnamefont{D.}~\bibnamefont{d'Enterria}} \bibnamefont{et~al.} (\bibinfo{year}{2019}), \eprint{1907.01435}.

\bibitem[{\citenamefont{Altarelli et~al.}(1998)}]{Altarelli:1998gn_alt}
\bibinfo{author}{\bibfnamefont{G.}~\bibnamefont{Altarelli}} \bibnamefont{et~al.}, \bibinfo{journal}{Nucl. Phys. B} \textbf{\bibinfo{volume}{534}}, \bibinfo{pages}{277} (\bibinfo{year}{1998}), \eprint{hep-ph/9806345}.

\bibitem[{\citenamefont{Candido et~al.}(2020)}]{Candido:2020yat_alt}
\bibinfo{author}{\bibfnamefont{A.}~\bibnamefont{Candido}} \bibnamefont{et~al.}, \bibinfo{journal}{JHEP} \textbf{\bibinfo{volume}{11}}, \bibinfo{pages}{129} (\bibinfo{year}{2020}), \eprint{2006.07377}.

\bibitem[{\citenamefont{Collins et~al.}(2022)}]{Collins:2021vke_alt}
\bibinfo{author}{\bibfnamefont{J.}~\bibnamefont{Collins}} \bibnamefont{et~al.}, \bibinfo{journal}{Phys. Rev. D} \textbf{\bibinfo{volume}{105}}, \bibinfo{pages}{076010} (\bibinfo{year}{2022}), \eprint{2111.01170}.

\bibitem[{\citenamefont{Candido et~al.}(2024)}]{Candido:2023ujx_alt}
\bibinfo{author}{\bibfnamefont{A.}~\bibnamefont{Candido}} \bibnamefont{et~al.}, \bibinfo{journal}{Eur. Phys. J. C} \textbf{\bibinfo{volume}{84}}, \bibinfo{pages}{335} (\bibinfo{year}{2024}), \eprint{2308.00025}.

\bibitem[{\citenamefont{Arroyo~Garcia et~al.}(2019)}]{Garcia:2019tne}
\bibinfo{author}{\bibfnamefont{A.}~\bibnamefont{Arroyo~Garcia}} \bibnamefont{et~al.}, \bibinfo{journal}{Phys. Lett. B} \textbf{\bibinfo{volume}{795}}, \bibinfo{pages}{569} (\bibinfo{year}{2019}), \eprint{1904.04394}.

\bibitem[{\citenamefont{Celiberto et~al.}(2018)}]{Celiberto:2018muu}
\bibinfo{author}{\bibfnamefont{F.~G.} \bibnamefont{Celiberto}} \bibnamefont{et~al.}, \bibinfo{journal}{Phys. Lett.} \textbf{\bibinfo{volume}{B786}}, \bibinfo{pages}{201} (\bibinfo{year}{2018}), \eprint{1808.09511}.

\bibitem[{\citenamefont{Bolognino et~al.}(2018)}]{Bolognino:2018rhb}
\bibinfo{author}{\bibfnamefont{A.~D.} \bibnamefont{Bolognino}} \bibnamefont{et~al.}, \bibinfo{journal}{Eur. Phys. J.} \textbf{\bibinfo{volume}{C78}}, \bibinfo{pages}{1023} (\bibinfo{year}{2018}), \eprint{1808.02395}.

\bibitem[{\citenamefont{Bolognino et~al.}(2021{\natexlab{a}})}]{Bolognino:2021niq}
\bibinfo{author}{\bibfnamefont{A.~D.} \bibnamefont{Bolognino}} \bibnamefont{et~al.}, \bibinfo{journal}{Eur. Phys. J. C} \textbf{\bibinfo{volume}{81}}, \bibinfo{pages}{846} (\bibinfo{year}{2021}{\natexlab{a}}), \eprint{2107.13415}.

\bibitem[{\citenamefont{Celiberto}(2019)}]{Celiberto:2019slj}
\bibinfo{author}{\bibfnamefont{F.~G.} \bibnamefont{Celiberto}}, \bibinfo{journal}{Nuovo Cim.} \textbf{\bibinfo{volume}{C42}}, \bibinfo{pages}{220} (\bibinfo{year}{2019}), \eprint{1912.11313}.

\bibitem[{\citenamefont{Silvetti\mbox{, M. Bonvini}}(2023)}]{Silvetti:2022hyc}
\bibinfo{author}{\bibfnamefont{F.}~\bibnamefont{Silvetti\mbox{, M. Bonvini}}}, \bibinfo{journal}{Eur. Phys. J. C} \textbf{\bibinfo{volume}{83}}, \bibinfo{pages}{267} (\bibinfo{year}{2023}), \eprint{2211.10142}.

\bibitem[{\citenamefont{Kang et~al.}(2024)\citenamefont{Kang, Li, and Salazar}}]{Kang:2023doo}
\bibinfo{author}{\bibfnamefont{Z.-B.} \bibnamefont{Kang}}, \bibinfo{author}{\bibfnamefont{E.}~\bibnamefont{Li}}, \bibnamefont{and} \bibinfo{author}{\bibfnamefont{F.}~\bibnamefont{Salazar}}, \bibinfo{journal}{JHEP} \textbf{\bibinfo{volume}{03}}, \bibinfo{pages}{027} (\bibinfo{year}{2024}), \eprint{2310.12102}.

\bibitem[{\citenamefont{Boer et~al.}(2016)}]{Boer:2015pni}
\bibinfo{author}{\bibfnamefont{D.}~\bibnamefont{Boer}} \bibnamefont{et~al.}, \bibinfo{journal}{Phys. Rev. Lett.} \textbf{\bibinfo{volume}{116}}, \bibinfo{pages}{122001} (\bibinfo{year}{2016}), \eprint{1511.03485}.

\bibitem[{\citenamefont{Lansberg et~al.}(2018)}]{Lansberg:2017dzg}
\bibinfo{author}{\bibfnamefont{J.-P.} \bibnamefont{Lansberg}} \bibnamefont{et~al.}, \bibinfo{journal}{Phys. Lett. B} \textbf{\bibinfo{volume}{784}}, \bibinfo{pages}{217} (\bibinfo{year}{2018}), \eprint{1710.01684}.

\bibitem[{\citenamefont{D'Alesio et~al.}(2020)}]{DAlesio:2020eqo}
\bibinfo{author}{\bibfnamefont{U.}~\bibnamefont{D'Alesio}} \bibnamefont{et~al.}, \bibinfo{journal}{Phys. Rev. D} \textbf{\bibinfo{volume}{102}}, \bibinfo{pages}{094011} (\bibinfo{year}{2020}), \eprint{2007.03353}.

\bibitem[{\citenamefont{Bacchetta et~al.}(2020)}]{Bacchetta:2020vty}
\bibinfo{author}{\bibfnamefont{A.}~\bibnamefont{Bacchetta}} \bibnamefont{et~al.}, \bibinfo{journal}{Eur. Phys. J. C} \textbf{\bibinfo{volume}{80}}, \bibinfo{pages}{733} (\bibinfo{year}{2020}), \eprint{2005.02288}.

\bibitem[{\citenamefont{Bacchetta et~al.}(2024)}]{Bacchetta:2024fci}
\bibinfo{author}{\bibfnamefont{A.}~\bibnamefont{Bacchetta}} \bibnamefont{et~al.}, \bibinfo{journal}{Eur. Phys. J. C} \textbf{\bibinfo{volume}{84}}, \bibinfo{pages}{576} (\bibinfo{year}{2024}), \eprint{2402.17556}.

\bibitem[{\citenamefont{Celiberto}(2021{\natexlab{a}})}]{Celiberto:2021zww}
\bibinfo{author}{\bibfnamefont{F.~G.} \bibnamefont{Celiberto}}, \bibinfo{journal}{Nuovo Cim.} \textbf{\bibinfo{volume}{C44}}, \bibinfo{pages}{36} (\bibinfo{year}{2021}{\natexlab{a}}), \eprint{2101.04630}.

\bibitem[{\citenamefont{Flore et~al.}(2020)}]{Flore:2020jau}
\bibinfo{author}{\bibfnamefont{C.}~\bibnamefont{Flore}} \bibnamefont{et~al.}, \bibinfo{journal}{Phys. Lett. B} \textbf{\bibinfo{volume}{811}}, \bibinfo{pages}{135926} (\bibinfo{year}{2020}), \eprint{2009.08264}.

\bibitem[{\citenamefont{Ball et~al.}(2022)}]{Ball:2022qks}
\bibinfo{author}{\bibfnamefont{R.~D.} \bibnamefont{Ball}} \bibnamefont{et~al.} (\bibinfo{collaboration}{NNPDF}), \bibinfo{journal}{Nature} \textbf{\bibinfo{volume}{608}}, \bibinfo{pages}{483} (\bibinfo{year}{2022}), \eprint{2208.08372}.

\bibitem[{\citenamefont{Guzzi et~al.}(2023)}]{Guzzi:2022rca}
\bibinfo{author}{\bibfnamefont{M.}~\bibnamefont{Guzzi}} \bibnamefont{et~al.}, \bibinfo{journal}{Phys. Lett. B} \textbf{\bibinfo{volume}{843}}, \bibinfo{pages}{137975} (\bibinfo{year}{2023}), \eprint{2211.01387}.

\bibitem[{\citenamefont{Ball et~al.}(2024{\natexlab{a}})}]{NNPDF:2023tyk_alt}
\bibinfo{author}{\bibfnamefont{R.~D.} \bibnamefont{Ball}} \bibnamefont{et~al.}, \bibinfo{journal}{Phys. Rev. D} \textbf{\bibinfo{volume}{109}}, \bibinfo{pages}{L091501} (\bibinfo{year}{2024}{\natexlab{a}}), \eprint{2311.00743}.

\bibitem[{\citenamefont{Caswell et~al.}(1986)}]{Caswell:1985ui}
\bibinfo{author}{\bibfnamefont{W.~E.} \bibnamefont{Caswell}} \bibnamefont{et~al.}, \bibinfo{journal}{Phys. Lett. B} \textbf{\bibinfo{volume}{167}}, \bibinfo{pages}{437} (\bibinfo{year}{1986}).

\bibitem[{\citenamefont{Bodwin et~al.}(1995)}]{Bodwin:1994jh}
\bibinfo{author}{\bibfnamefont{G.~T.} \bibnamefont{Bodwin}} \bibnamefont{et~al.}, \bibinfo{journal}{Phys. Rev. D} \textbf{\bibinfo{volume}{51}}, \bibinfo{pages}{1125} (\bibinfo{year}{1995}), \eprint{hep-ph/9407339}.

\bibitem[{\citenamefont{Mele\mbox{, P. Nason}}(1991)}]{Mele:1990cw}
\bibinfo{author}{\bibfnamefont{B.}~\bibnamefont{Mele\mbox{, P. Nason}}}, \bibinfo{journal}{Nucl. Phys. B} \textbf{\bibinfo{volume}{361}}, \bibinfo{pages}{626} (\bibinfo{year}{1991}).

\bibitem[{\citenamefont{Cacciari et~al.}(1994{\natexlab{a}})}]{Cacciari:1993mq_alt}
\bibinfo{author}{\bibfnamefont{M.}~\bibnamefont{Cacciari}} \bibnamefont{et~al.}, \bibinfo{journal}{Nucl. Phys. B} \textbf{\bibinfo{volume}{421}}, \bibinfo{pages}{530} (\bibinfo{year}{1994}{\natexlab{a}}), \eprint{hep-ph/9311260}.

\bibitem[{\citenamefont{Cacciari et~al.}(2012)}]{Cacciari:2012ny}
\bibinfo{author}{\bibfnamefont{M.}~\bibnamefont{Cacciari}} \bibnamefont{et~al.}, \bibinfo{journal}{JHEP} \textbf{\bibinfo{volume}{10}}, \bibinfo{pages}{137} (\bibinfo{year}{2012}), \eprint{1205.6344}.

\bibitem[{\citenamefont{Helenius\mbox{, H. Paukkunen}}(2018)}]{Helenius:2018uul}
\bibinfo{author}{\bibfnamefont{I.}~\bibnamefont{Helenius\mbox{, H. Paukkunen}}}, \bibinfo{journal}{JHEP} \textbf{\bibinfo{volume}{05}}, \bibinfo{pages}{196} (\bibinfo{year}{2018}), \eprint{1804.03557}.

\bibitem[{\citenamefont{Helenius\mbox{, H. Paukkunen}}(2023)}]{Helenius:2023wkn}
\bibinfo{author}{\bibfnamefont{I.}~\bibnamefont{Helenius\mbox{, H. Paukkunen}}}, \bibinfo{journal}{JHEP} \textbf{\bibinfo{volume}{07}}, \bibinfo{pages}{054} (\bibinfo{year}{2023}), \eprint{2303.17864}.

\bibitem[{\citenamefont{Czakon et~al.}(2021)}]{Czakon:2021ohs}
\bibinfo{author}{\bibfnamefont{M.}~\bibnamefont{Czakon}} \bibnamefont{et~al.}, \bibinfo{journal}{JHEP} \textbf{\bibinfo{volume}{10}}, \bibinfo{pages}{216} (\bibinfo{year}{2021}), \eprint{2102.08267}.

\bibitem[{\citenamefont{Czakon et~al.}(2023)}]{Czakon:2022pyz}
\bibinfo{author}{\bibfnamefont{M.}~\bibnamefont{Czakon}} \bibnamefont{et~al.}, \bibinfo{journal}{JHEP} \textbf{\bibinfo{volume}{03}}, \bibinfo{pages}{251} (\bibinfo{year}{2023}), \eprint{2210.06078}.

\bibitem[{\citenamefont{Generet}(2023)}]{Generet:2023vte}
\bibinfo{author}{\bibfnamefont{T.}~\bibnamefont{Generet}}, Ph.D. thesis, \bibinfo{school}{RWTH Aachen University} (\bibinfo{year}{2023}).

\bibitem[{\citenamefont{Ghira et~al.}(2023)}]{Ghira:2023bxr_alt}
\bibinfo{author}{\bibfnamefont{A.}~\bibnamefont{Ghira}} \bibnamefont{et~al.}, \bibinfo{journal}{JHEP} \textbf{\bibinfo{volume}{11}}, \bibinfo{pages}{120} (\bibinfo{year}{2023}), \eprint{2309.06139}.

\bibitem[{\citenamefont{Bonino et~al.}(2024)}]{Bonino:2023icn_alt}
\bibinfo{author}{\bibfnamefont{L.}~\bibnamefont{Bonino}} \bibnamefont{et~al.}, \bibinfo{journal}{JHEP} \textbf{\bibinfo{volume}{06}}, \bibinfo{pages}{040} (\bibinfo{year}{2024}), \eprint{2312.12519}.

\bibitem[{\citenamefont{Cacciari et~al.}(2024)}]{Cacciari:2024kaa}
\bibinfo{author}{\bibfnamefont{M.}~\bibnamefont{Cacciari}} \bibnamefont{et~al.}, \bibinfo{journal}{Eur. Phys. J. C} \textbf{\bibinfo{volume}{84}}, \bibinfo{pages}{889} (\bibinfo{year}{2024}), \eprint{2406.04173}.

\bibitem[{\citenamefont{Cacciari et~al.}(1994{\natexlab{b}})}]{Cacciari:1994dr_alt}
\bibinfo{author}{\bibfnamefont{M.}~\bibnamefont{Cacciari}} \bibnamefont{et~al.}, \bibinfo{journal}{Phys. Rev. Lett.} \textbf{\bibinfo{volume}{73}}, \bibinfo{pages}{1586} (\bibinfo{year}{1994}{\natexlab{b}}), \eprint{hep-ph/9405241}.

\bibitem[{\citenamefont{Celiberto}(2024{\natexlab{a}})}]{Celiberto:2024mex_article}
\bibinfo{author}{\bibfnamefont{F.~G.} \bibnamefont{Celiberto}}, \bibinfo{journal}{\emph{Proceedings of Moriond QCD}}  (\bibinfo{year}{2024}{\natexlab{a}}), \eprint{2405.08221}.

\bibitem[{\citenamefont{Celiberto}(2024{\natexlab{b}})}]{Celiberto:2024bxu}
\bibinfo{author}{\bibfnamefont{F.~G.} \bibnamefont{Celiberto}}, \bibinfo{journal}{PoS} \textbf{\bibinfo{volume}{DIS2024}}, \bibinfo{pages}{168} (\bibinfo{year}{2024}{\natexlab{b}}), \eprint{2406.10779}.

\bibitem[{\citenamefont{Forte et~al.}(2002)}]{Forte:2002fg}
\bibinfo{author}{\bibfnamefont{S.}~\bibnamefont{Forte}} \bibnamefont{et~al.}, \bibinfo{journal}{JHEP} \textbf{\bibinfo{volume}{05}}, \bibinfo{pages}{062} (\bibinfo{year}{2002}), \eprint{hep-ph/0204232}.

\bibitem[{\citenamefont{Kang et~al.}(2014)}]{Kang:2014tta}
\bibinfo{author}{\bibfnamefont{Z.-B.} \bibnamefont{Kang}} \bibnamefont{et~al.}, \bibinfo{journal}{Phys. Rev. D} \textbf{\bibinfo{volume}{90}}, \bibinfo{pages}{034006} (\bibinfo{year}{2014}), \eprint{1401.0923}.

\bibitem[{\citenamefont{Echevarria}(2019)}]{Echevarria:2019ynx}
\bibinfo{author}{\bibfnamefont{M.~G.} \bibnamefont{Echevarria}}, \bibinfo{journal}{JHEP} \textbf{\bibinfo{volume}{10}}, \bibinfo{pages}{144} (\bibinfo{year}{2019}), \eprint{1907.06494}.

\bibitem[{\citenamefont{Boer et~al.}(2023)}]{Boer:2023zit}
\bibinfo{author}{\bibfnamefont{D.}~\bibnamefont{Boer}} \bibnamefont{et~al.}, \bibinfo{journal}{JHEP} \textbf{\bibinfo{volume}{08}}, \bibinfo{pages}{105} (\bibinfo{year}{2023}), \eprint{2304.09473}.

\bibitem[{\citenamefont{Celiberto}(2017)}]{Celiberto:2017ius}
\bibinfo{author}{\bibfnamefont{F.~G.} \bibnamefont{Celiberto}}, Ph.D. thesis, \bibinfo{school}{U. Calabria \& INFN} (\bibinfo{year}{2017}), \eprint{1707.04315}.

\bibitem[{\citenamefont{Celiberto}(2021{\natexlab{b}})}]{Celiberto:2020wpk}
\bibinfo{author}{\bibfnamefont{F.~G.} \bibnamefont{Celiberto}}, \bibinfo{journal}{Eur. Phys. J. C} \textbf{\bibinfo{volume}{81}}, \bibinfo{pages}{691} (\bibinfo{year}{2021}{\natexlab{b}}), \eprint{2008.07378}.

\bibitem[{\citenamefont{Celiberto}(2022{\natexlab{a}})}]{Celiberto:2022rfj}
\bibinfo{author}{\bibfnamefont{F.~G.} \bibnamefont{Celiberto}}, \bibinfo{journal}{Phys. Rev. D} \textbf{\bibinfo{volume}{105}}, \bibinfo{pages}{114008} (\bibinfo{year}{2022}{\natexlab{a}}), \eprint{2204.06497}.

\bibitem[{\citenamefont{Celiberto}(2023{\natexlab{a}})}]{Celiberto:2022kxx}
\bibinfo{author}{\bibfnamefont{F.~G.} \bibnamefont{Celiberto}}, \bibinfo{journal}{Eur. Phys. J. C} \textbf{\bibinfo{volume}{83}}, \bibinfo{pages}{332} (\bibinfo{year}{2023}{\natexlab{a}}), \eprint{2208.14577}.

\bibitem[{\citenamefont{Celiberto et~al.}(2021{\natexlab{a}})}]{Celiberto:2020tmb}
\bibinfo{author}{\bibfnamefont{F.~G.} \bibnamefont{Celiberto}} \bibnamefont{et~al.}, \bibinfo{journal}{Eur. Phys. J. C} \textbf{\bibinfo{volume}{81}}, \bibinfo{pages}{293} (\bibinfo{year}{2021}{\natexlab{a}}), \eprint{2008.00501}.

\bibitem[{\citenamefont{Bolognino et~al.}(2021{\natexlab{b}})}]{Bolognino:2021mrc}
\bibinfo{author}{\bibfnamefont{A.~D.} \bibnamefont{Bolognino}} \bibnamefont{et~al.}, \bibinfo{journal}{Phys. Rev. D} \textbf{\bibinfo{volume}{103}}, \bibinfo{pages}{094004} (\bibinfo{year}{2021}{\natexlab{b}}), \eprint{2103.07396}.

\bibitem[{\citenamefont{Celiberto et~al.}(2021{\natexlab{b}})}]{Celiberto:2021dzy}
\bibinfo{author}{\bibfnamefont{F.~G.} \bibnamefont{Celiberto}} \bibnamefont{et~al.}, \bibinfo{journal}{Eur. Phys. J. C} \textbf{\bibinfo{volume}{81}}, \bibinfo{pages}{780} (\bibinfo{year}{2021}{\natexlab{b}}), \eprint{2105.06432}.

\bibitem[{\citenamefont{Celiberto et~al.}(2021{\natexlab{c}})}]{Celiberto:2021fdp}
\bibinfo{author}{\bibfnamefont{F.~G.} \bibnamefont{Celiberto}} \bibnamefont{et~al.}, \bibinfo{journal}{Phys. Rev. D} \textbf{\bibinfo{volume}{104}}, \bibinfo{pages}{114007} (\bibinfo{year}{2021}{\natexlab{c}}), \eprint{2109.11875}.

\bibitem[{\citenamefont{Celiberto et~al.}(2022{\natexlab{a}})}]{Celiberto:2022zdg}
\bibinfo{author}{\bibfnamefont{F.~G.} \bibnamefont{Celiberto}} \bibnamefont{et~al.}, \bibinfo{journal}{Phys. Rev. D} \textbf{\bibinfo{volume}{105}}, \bibinfo{pages}{114056} (\bibinfo{year}{2022}{\natexlab{a}}), \eprint{2205.13429}.

\bibitem[{\citenamefont{Celiberto et~al.}(2022{\natexlab{b}})}]{Celiberto:2022gji}
\bibinfo{author}{\bibfnamefont{F.~G.} \bibnamefont{Celiberto}} \bibnamefont{et~al.}, \bibinfo{journal}{Phys. Rev. D} \textbf{\bibinfo{volume}{106}}, \bibinfo{pages}{114004} (\bibinfo{year}{2022}{\natexlab{b}}), \eprint{2207.05015}.

\bibitem[{\citenamefont{Harland-Lang et~al.}(2019)}]{Harland-Lang:2018bxd_alt}
\bibinfo{author}{\bibfnamefont{L.~A.} \bibnamefont{Harland-Lang}} \bibnamefont{et~al.}, \bibinfo{journal}{Eur. Phys. J. C} \textbf{\bibinfo{volume}{79}}, \bibinfo{pages}{225} (\bibinfo{year}{2019}), \eprint{1811.08434}.

\bibitem[{\citenamefont{Ball et~al.}(2024{\natexlab{b}})}]{NNPDF:2024dpb}
\bibinfo{author}{\bibfnamefont{R.~D.} \bibnamefont{Ball}} \bibnamefont{et~al.} (\bibinfo{collaboration}{NNPDF}), \bibinfo{journal}{Eur. Phys. J. C} \textbf{\bibinfo{volume}{84}}, \bibinfo{pages}{517} (\bibinfo{year}{2024}{\natexlab{b}}), \eprint{2401.10319}.

\bibitem[{\citenamefont{Kassabov et~al.}(2023)}]{Kassabov:2022orn_alt}
\bibinfo{author}{\bibfnamefont{Z.}~\bibnamefont{Kassabov}} \bibnamefont{et~al.}, \bibinfo{journal}{JHEP} \textbf{\bibinfo{volume}{03}}, \bibinfo{pages}{148} (\bibinfo{year}{2023}), \eprint{2207.07616}.

\bibitem[{\citenamefont{Altheimer et~al.}(2014)}]{Altheimer:2013yza}
\bibinfo{author}{\bibfnamefont{A.}~\bibnamefont{Altheimer}} \bibnamefont{et~al.}, \bibinfo{journal}{Eur. Phys. J. C} \textbf{\bibinfo{volume}{74}}, \bibinfo{pages}{2792} (\bibinfo{year}{2014}), \eprint{1311.2708}.

\bibitem[{\citenamefont{Adams et~al.}(2015)}]{Adams:2015hiv}
\bibinfo{author}{\bibfnamefont{D.}~\bibnamefont{Adams}} \bibnamefont{et~al.}, \bibinfo{journal}{Eur. Phys. J. C} \textbf{\bibinfo{volume}{75}}, \bibinfo{pages}{409} (\bibinfo{year}{2015}), \eprint{1504.00679}.

\bibitem[{\citenamefont{Procura et~al.}(2010)}]{Procura:2009vm_alt}
\bibinfo{author}{\bibfnamefont{M.}~\bibnamefont{Procura}} \bibnamefont{et~al.}, \bibinfo{journal}{Phys. Rev. D} \textbf{\bibinfo{volume}{81}}, \bibinfo{pages}{074009} (\bibinfo{year}{2010}), \eprint{0911.4980}.

\bibitem[{\citenamefont{Bauer and Mereghetti}(2014)}]{Bauer:2013bza}
\bibinfo{author}{\bibfnamefont{C.~W.} \bibnamefont{Bauer}} \bibnamefont{and} \bibinfo{author}{\bibfnamefont{E.}~\bibnamefont{Mereghetti}}, \bibinfo{journal}{JHEP} \textbf{\bibinfo{volume}{04}}, \bibinfo{pages}{051} (\bibinfo{year}{2014}), \eprint{1312.5605}.

\bibitem[{\citenamefont{Chien et~al.}(2016)}]{Chien:2015ctp}
\bibinfo{author}{\bibfnamefont{Y.-T.} \bibnamefont{Chien}} \bibnamefont{et~al.}, \bibinfo{journal}{JHEP} \textbf{\bibinfo{volume}{05}}, \bibinfo{pages}{125} (\bibinfo{year}{2016}), \eprint{1512.06851}.

\bibitem[{\citenamefont{Maltoni et~al.}(2016)}]{Maltoni:2016ays_alt}
\bibinfo{author}{\bibfnamefont{F.}~\bibnamefont{Maltoni}} \bibnamefont{et~al.}, \bibinfo{journal}{Phys. Rev. D} \textbf{\bibinfo{volume}{94}}, \bibinfo{pages}{054015} (\bibinfo{year}{2016}), \eprint{1606.03449}.

\bibitem[{\citenamefont{Kang et~al.}(2017{\natexlab{a}})}]{Kang:2017glf}
\bibinfo{author}{\bibfnamefont{Z.-B.} \bibnamefont{Kang}} \bibnamefont{et~al.}, \bibinfo{journal}{JHEP} \textbf{\bibinfo{volume}{11}}, \bibinfo{pages}{068} (\bibinfo{year}{2017}{\natexlab{a}}), \eprint{1705.08443}.

\bibitem[{\citenamefont{Metodiev et~al.}(2018)}]{Metodiev:2018ftz_alt}
\bibinfo{author}{\bibfnamefont{E.}~\bibnamefont{Metodiev}} \bibnamefont{et~al.}, \bibinfo{journal}{Phys. Rev. Lett.} \textbf{\bibinfo{volume}{120}}, \bibinfo{pages}{241602} (\bibinfo{year}{2018}), \eprint{1802.00008}.

\bibitem[{\citenamefont{Marzani et~al.}()\citenamefont{Marzani, Soyez, and Spannowsky}}]{Marzani:2019hun_article}
\bibinfo{author}{\bibfnamefont{S.}~\bibnamefont{Marzani}}, \bibinfo{author}{\bibfnamefont{G.}~\bibnamefont{Soyez}}, \bibnamefont{and} \bibinfo{author}{\bibfnamefont{M.}~\bibnamefont{Spannowsky}} (????), \eprint{1901.10342}.

\bibitem[{\citenamefont{Kasieczka et~al.}(2020)}]{Kasieczka:2020nyd}
\bibinfo{author}{\bibfnamefont{G.}~\bibnamefont{Kasieczka}} \bibnamefont{et~al.}, \bibinfo{journal}{JHEP} \textbf{\bibinfo{volume}{09}}, \bibinfo{pages}{195} (\bibinfo{year}{2020}), \eprint{2007.04319}.

\bibitem[{\citenamefont{Nachman et~al.}(2022)}]{Nachman:2022emq}
\bibinfo{author}{\bibfnamefont{B.}~\bibnamefont{Nachman}} \bibnamefont{et~al.}, \bibinfo{journal}{Front. in Phys.} \textbf{\bibinfo{volume}{10}}, \bibinfo{pages}{897719} (\bibinfo{year}{2022}), \eprint{2203.07462}.

\bibitem[{\citenamefont{Dhani et~al.}(2024)}]{Dhani:2024gtx}
\bibinfo{author}{\bibfnamefont{P.~K.} \bibnamefont{Dhani}} \bibnamefont{et~al.} (\bibinfo{year}{2024}), \eprint{2410.05415}.

\bibitem[{\citenamefont{Kang et~al.}(2017{\natexlab{b}})}]{Kang:2017yde}
\bibinfo{author}{\bibfnamefont{Z.-B.} \bibnamefont{Kang}} \bibnamefont{et~al.}, \bibinfo{journal}{Phys. Rev. Lett.} \textbf{\bibinfo{volume}{119}}, \bibinfo{pages}{032001} (\bibinfo{year}{2017}{\natexlab{b}}), \eprint{1702.03287}.

\bibitem[{\citenamefont{Baumgart et~al.}(2014)}]{Baumgart:2014upa}
\bibinfo{author}{\bibfnamefont{M.}~\bibnamefont{Baumgart}} \bibnamefont{et~al.}, \bibinfo{journal}{JHEP} \textbf{\bibinfo{volume}{11}}, \bibinfo{pages}{003} (\bibinfo{year}{2014}), \eprint{1406.2295}.

\bibitem[{\citenamefont{Kang et~al.}(2016)}]{Kang:2016ehg_alt}
\bibinfo{author}{\bibfnamefont{Z.-B.} \bibnamefont{Kang}} \bibnamefont{et~al.}, \bibinfo{journal}{JHEP} \textbf{\bibinfo{volume}{11}}, \bibinfo{pages}{155} (\bibinfo{year}{2016}), \eprint{1606.07063}.

\bibitem[{\citenamefont{Luisoni\mbox{, S. Marzani}}(2015)}]{Luisoni:2015xha}
\bibinfo{author}{\bibfnamefont{G.}~\bibnamefont{Luisoni\mbox{, S. Marzani}}}, \bibinfo{journal}{J. Phys. G} \textbf{\bibinfo{volume}{42}}, \bibinfo{pages}{103101} (\bibinfo{year}{2015}), \eprint{1505.04084}.

\bibitem[{\citenamefont{Caletti et~al.}(2021)}]{Caletti:2021oor}
\bibinfo{author}{\bibfnamefont{S.}~\bibnamefont{Caletti}} \bibnamefont{et~al.}, \bibinfo{journal}{JHEP} \textbf{\bibinfo{volume}{07}}, \bibinfo{pages}{076} (\bibinfo{year}{2021}), \eprint{2104.06920}.

\bibitem[{\citenamefont{Reichelt et~al.}(2022)}]{Reichelt:2021svh}
\bibinfo{author}{\bibfnamefont{D.}~\bibnamefont{Reichelt}} \bibnamefont{et~al.}, \bibinfo{journal}{JHEP} \textbf{\bibinfo{volume}{03}}, \bibinfo{pages}{131} (\bibinfo{year}{2022}), \eprint{2112.09545}.

\bibitem[{\citenamefont{Celiberto et~al.}(2022{\natexlab{c}})}]{Celiberto:2022dyf_alt}
\bibinfo{author}{\bibfnamefont{F.~G.} \bibnamefont{Celiberto}} \bibnamefont{et~al.}, \bibinfo{journal}{Eur. Phys. J. C} \textbf{\bibinfo{volume}{82}}, \bibinfo{pages}{929} (\bibinfo{year}{2022}{\natexlab{c}}), \eprint{2202.12227}.

\bibitem[{\citenamefont{Celiberto}(2023{\natexlab{b}})}]{Celiberto:2023fzz}
\bibinfo{author}{\bibfnamefont{F.~G.} \bibnamefont{Celiberto}}, \bibinfo{journal}{Universe} \textbf{\bibinfo{volume}{9}}, \bibinfo{pages}{324} (\bibinfo{year}{2023}{\natexlab{b}}), \eprint{2305.14295}.

\bibitem[{\citenamefont{Celiberto}(2022{\natexlab{b}})}]{Celiberto:2022keu}
\bibinfo{author}{\bibfnamefont{F.~G.} \bibnamefont{Celiberto}}, \bibinfo{journal}{Phys. Lett. B} \textbf{\bibinfo{volume}{835}}, \bibinfo{pages}{137554} (\bibinfo{year}{2022}{\natexlab{b}}), \eprint{2206.09413}.

\bibitem[{\citenamefont{Celiberto}(2024{\natexlab{c}})}]{Celiberto:2024omj}
\bibinfo{author}{\bibfnamefont{F.~G.} \bibnamefont{Celiberto}}, \bibinfo{journal}{Eur. Phys. J. C} \textbf{\bibinfo{volume}{84}}, \bibinfo{pages}{384} (\bibinfo{year}{2024}{\natexlab{c}}), \eprint{2401.01410}.

\bibitem[{\citenamefont{Chapon et~al.}(2022)}]{Chapon:2020heu_alt}
\bibinfo{author}{\bibfnamefont{E.}~\bibnamefont{Chapon}} \bibnamefont{et~al.}, \bibinfo{journal}{Prog. Part. Nucl. Phys.} \textbf{\bibinfo{volume}{122}}, \bibinfo{pages}{103906} (\bibinfo{year}{2022}).

\bibitem[{\citenamefont{Amoroso et~al.}(2022)}]{Amoroso:2022eow}
\bibinfo{author}{\bibfnamefont{S.}~\bibnamefont{Amoroso}} \bibnamefont{et~al.}, \bibinfo{journal}{Acta Phys. Polon. B} \textbf{\bibinfo{volume}{53}}, \bibinfo{pages}{A1} (\bibinfo{year}{2022}), \eprint{2203.13923}.

\bibitem[{\citenamefont{A.~Khalek et~al.}(2022{\natexlab{a}})}]{AbdulKhalek:2021gbh}
\bibinfo{author}{\bibfnamefont{R.}~\bibnamefont{A.~Khalek}} \bibnamefont{et~al.}, \bibinfo{journal}{Nucl. Phys. A} \textbf{\bibinfo{volume}{1026}}, \bibinfo{pages}{122447} (\bibinfo{year}{2022}{\natexlab{a}}), \eprint{2103.05419}.

\bibitem[{\citenamefont{A.~Khalek et~al.}(2022{\natexlab{b}})}]{Khalek:2022bzd}
\bibinfo{author}{\bibfnamefont{R.}~\bibnamefont{A.~Khalek}} \bibnamefont{et~al.} (\bibinfo{year}{2022}{\natexlab{b}}), \eprint{2203.13199}.

\bibitem[{\citenamefont{Abir et~al.}(2023)}]{Abir:2023fpo}
\bibinfo{author}{\bibfnamefont{R.}~\bibnamefont{Abir}} \bibnamefont{et~al.} (\bibinfo{year}{2023}), \eprint{2305.14572}.

\bibitem[{\citenamefont{Adachi et~al.}(2022)}]{AlexanderAryshev:2022pkx}
\bibinfo{author}{\bibfnamefont{I.}~\bibnamefont{Adachi}} \bibnamefont{et~al.} (\bibinfo{collaboration}{ILC International Community}) (\bibinfo{year}{2022}), \eprint{2203.07622}.

\bibitem[{\citenamefont{Allaire et~al.}(2024)}]{Allaire:2023fgp}
\bibinfo{author}{\bibfnamefont{C.}~\bibnamefont{Allaire}} \bibnamefont{et~al.}, \bibinfo{journal}{Comput. Softw. Big Sci.} \textbf{\bibinfo{volume}{8}}, \bibinfo{pages}{5} (\bibinfo{year}{2024}), \eprint{2307.08593}.

\bibitem[{\citenamefont{Hekhorn}(2024)}]{Hekhorn:2024jrj_article}
\bibinfo{author}{\bibfnamefont{F.}~\bibnamefont{Hekhorn}}, \bibinfo{journal}{\emph{Proceedings of DIS}}  (\bibinfo{year}{2024}), \eprint{2406.06083}.

\bibitem[{\citenamefont{Hammou et~al.}(2023)}]{Hammou:2023heg}
\bibinfo{author}{\bibfnamefont{E.}~\bibnamefont{Hammou}} \bibnamefont{et~al.}, \bibinfo{journal}{JHEP} \textbf{\bibinfo{volume}{11}}, \bibinfo{pages}{090} (\bibinfo{year}{2023}), \eprint{2307.10370}.

\bibitem[{\citenamefont{Costantini et~al.}(2024)}]{Costantini:2024xae}
\bibinfo{author}{\bibfnamefont{M.~N.} \bibnamefont{Costantini}} \bibnamefont{et~al.} (\bibinfo{year}{2024}), \eprint{2402.03308}.

\bibitem[{\citenamefont{Hammou}(2024)}]{Hammou:2024cwu_article}
\bibinfo{author}{\bibfnamefont{E.}~\bibnamefont{Hammou}}, \bibinfo{journal}{\emph{Proceedings of Moriond QCD}}  (\bibinfo{year}{2024}), \eprint{2405.09270}.

\bibitem[{\citenamefont{Celiberto et~al.}(2024{\natexlab{a}})}]{Celiberto:2023rzw_alt}
\bibinfo{author}{\bibfnamefont{F.~G.} \bibnamefont{Celiberto}} \bibnamefont{et~al.}, \bibinfo{journal}{Phys. Lett. B} \textbf{\bibinfo{volume}{848}}, \bibinfo{pages}{138406} (\bibinfo{year}{2024}{\natexlab{a}}), \eprint{2308.00809}.

\bibitem[{\citenamefont{Celiberto}(2024{\natexlab{d}})}]{Celiberto:2024mrq}
\bibinfo{author}{\bibfnamefont{F.~G.} \bibnamefont{Celiberto}}, \bibinfo{journal}{Symmetry} \textbf{\bibinfo{volume}{16}}, \bibinfo{pages}{550} (\bibinfo{year}{2024}{\natexlab{d}}), \eprint{2403.15639}.

\bibitem[{\citenamefont{Celiberto et~al.}(2024{\natexlab{b}})}]{Celiberto:2024mab_alt}
\bibinfo{author}{\bibfnamefont{F.~G.} \bibnamefont{Celiberto}} \bibnamefont{et~al.}, \bibinfo{journal}{Eur. Phys. J. C} \textbf{\bibinfo{volume}{84}}, \bibinfo{pages}{1071} (\bibinfo{year}{2024}{\natexlab{b}}), \eprint{2405.14773}.

\end{thebibliography}

\end{document}